\documentclass[10pt, letter]{IEEEtran}
\usepackage[dvips]{graphicx}
\usepackage{subfigure}
\usepackage{times}
\usepackage{latexsym}
\usepackage{amsmath}
\usepackage{amssymb}
\usepackage{multirow}
\usepackage{algorithmic}
\usepackage{eucal}
\usepackage{epsfig}
\usepackage{cite}
\usepackage{longtable}
\usepackage{slashbox}
\usepackage{amsthm}
\usepackage{latexsym,bm}
\usepackage{color}
\usepackage{subfigure}

\newcommand{\mW}{\bm{W}}

\theoremstyle{plain}
\newtheorem{theorem}{Theorem}

\theoremstyle{definition}
\theoremstyle{definition}

\begin{document}

\title{Performance of MIMO Relay DCSK-CD Systems over Nakagami Fading Channels
\thanks{This work was supported by the NSF of China (Nos. 60972053 and 61001073), the European Union-FP7 (CoNHealth, No. 294923),
as well as the Hong Kong Research Grants Council (No. CityU1114/11E).}
\thanks{Y. Fang, Jing Xu and L. Wang are with the Department of Communication Engineering,
 Xiamen University, Fujian, 361005, China (email: fangyi1986812@163.com, 491592946@qq.com, wanglin@xmu.edu.cn)}
\thanks{Guanrong Chen is with with the Department of Electronic Engineering,
City University of Hong Kong, Hong Kong SAR, China (email: gchen@ee.cityu.edu.hk)}
\thanks{Copyright (c) 2012 IEEE. Personal use of this material is permitted.
However, permission to use this material for any other purposes must be
obtained from the IEEE by sending an email to pubs-permissions@ieee.org.
}}

\author{Yi Fang, Jing Xu, Lin Wang, \emph{Senior Member, IEEE}, and Guanrong Chen, \emph{Fellow, IEEE}}

\maketitle
\begin{abstract}
A multi-access multiple-input multiple-output (MIMO) relay differential chaos shift keying cooperative diversity (DCSK-CD) system is proposed in this paper as a comprehensive cooperation scheme, in which the relay and destination both employ multiple antennas to strengthen the robustness against signal fading in a wireless network. It is shown that, with spatial diversity gains, the bit error rate (BER) performance of the proposed system is remarkably better than the conventional DCSK non-cooperation (DCSK-NC) and DCSK cooperative communication (DCSK-CC) systems. Moreover, the exact BER and close-form expressions of the proposed system are derived over Nakagami fading channels through the moment generating function (MGF), which is shown to be highly consistent with the simulation results. Meanwhile, this paper illustrates a trade-off between the performance and the complexity, and provides a threshold for the number of relay antennas keeping the user consumed energy constant. Due to the above-mentioned advantages, the proposed system stands out as a good candidate or alternative for energy-constrained wireless communications based on chaotic modulation, especially for low-power and low-cost wireless personal area networks (WPANs).
\end{abstract}



\section{Introduction}
Inspired by the wideband nature, chaotic signals have shown a great promising for spread-spectrum communication systems \cite{Kolumbam96,899917,1015003}. Among the digital communication schemes proposed so far, the chaos shift keying (CSK) and differential CSK (DCSK) have been widely studied and analyzed in recent years \cite{lau2003chaos,899922,yelifen05,5200887,Kaddoum2010} attracted by their excellent capacity against multipath fading and time delay. Between these two schemes, DCSK is a non-coherent method which only requires frame or symbol rate sampling instead of channel estimation \cite{675115}. Furthermore, the hardware advantage of the DCSK scheme does not demand extra spreading and dispreading circuits in the mono-user case \cite{246164}. As a result, this method is more practical at the expense of some performance degrading in comparison with its coherent counterpart.

In a wireless network, signal fading aroused from multipath propagation is one important factor that influences the system performance. Many methods have been proposed to overcome this weakness. On the one hand, some chaotic cooperative communication systems have been suggested to deal with multipath fading, e.g., the DCSK cooperative communication (DCSK-CC) systems \cite{5537139,5629387}, which can provide a considerable gain over the conventional DCSK systems. On the other hand, multi-antenna techniques are particularly attractive as they can be easily combined with other forms of diversity. The use of multiple antennas at all terminals or only the receiver ends of a wireless link, which are called multiple-input multiple-output (MIMO) or single-input multiple-output (SIMO) techniques, demonstrates a significant improvement in link reliability and spectral efficiency \cite{1194444,983319}. Motivated by these observations of advantages, several MIMO \cite{5648457,5937879} and SIMO systems \cite{4469995} have been carefully studied aiming to improve the performance of the DCSK communications over fading channels. As a potential application, applying MIMO technique to relay networks has been considered in recent years \cite{4202183, 1336724}, referred to as cooperative diversity. Moreover, some theoretical research on MIMO relay channels has been carried out, such as their capacity analysis \cite{1638664,1377490}.

Due to the advantages of low cost and low complexity, some ultra-wideband (UWB) systems based on DCSK or FM-DCSK have been considered for wireless personal area networks (WPANs) and wireless local networks (WLANs) \cite{5383640,5371909,Kis2001}. In addition, a SIMO FM-DCSK UWB system has been proposed \cite{5754623} to further improve the system performance in such environments.

However, little research has been conducted focusing on the performance improvement of the multi-access DCSK systems. To address this problem, the MIMO relay technique is introduced to the multi-access DCSK system in the present paper, which will be called an MIMO relay DCSK-cooperation diversity (DCSK-CD) system. Although channels are assumed to be Rayleigh fading in most studies for DCSK systems,
it was pointed out that path fading statistics are more suitably to be described by a Nakagami distribution \cite{380145}. In fact, Nakagami channels have already been used for research on cooperation systems \cite{4573263,5184915}. Motivated by this observation, the presently proposed system will be investigated over fading channels subjected to Nakagami distribution.
In the proposed system, the relay and destination can employ multiple antennas to construct two SIMO and one MIMO systems of DCSK, on the source-relay link, source-destination link and relay-destination link, respectively. The new cooperation scheme can significantly enhance the capacity against signal fading in a wireless multipath fading channel with moderate complexity.
 The performance of the proposed system is much better than the conventional DCSK-NC and DCSK-CC ones \cite{5629387} according to our theoretical analyses and simulated results, to be reported below, where more gain can be achieved as the number of antennas used increases. Moreover, a threshold for the number of relay's antennas will be derived provided that each user is subjected to an energy constraint. Furthermore, a close-form BER expression over Nakagami fading channels will be derived by using moment generating function (MGF), which agrees well with the numerical simulations.

The remainder of this paper is organized as follows. In Section~\ref{sect:system}, the principles of DCSK modulation and the MIMO relay DCSK-CD system are reviewed and described, respectively. In Section~\ref{sect:Performance}, performance analysis on the new system is carried out, with a close form BER expression derived. Simulation results are shown in Section~\ref{sect:SIMU} and conclusions are given in Section~\ref{sect:Conclusions}.

\section{System Model}
\label{sect:system}
In this section, the multi-user DCSK system is introduced based on Walsh code, and then a new multi-access MIMO relay DCSK-CD system is proposed. In the rest of this paper, it is assumed that all the independent channels satisfy the same channel condition, that is, all channels have the same Gaussian-distributed random noise with the power spectral density of $N_0$. Also, equal-gain combiner (EGC) is adopted at the receivers in the systems because it not only can improve the error performance but also can be easily implemented \cite{558680}.

\subsection{Principle of DCSK Modulation} \label{sect:conv_DCSK}
\begin{figure}[tbp]
\centering
\subfigure[]{ \label{fig:subfig:a} 
\includegraphics[width=3.2in,height=0.65in]{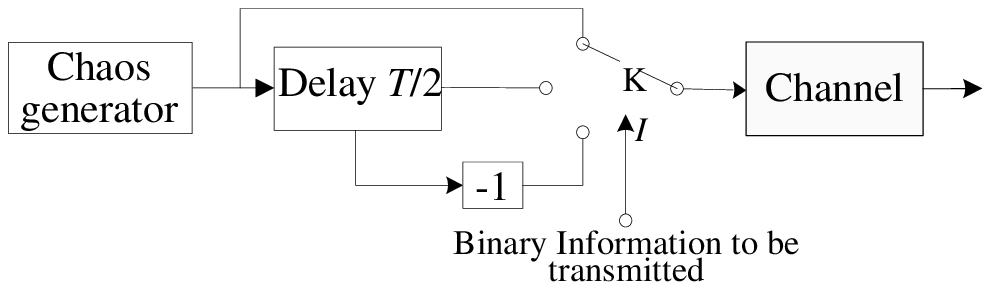}}
\subfigure[]{ \label{fig:subfig:b} 
\includegraphics[width=3.2in,height=0.5in]{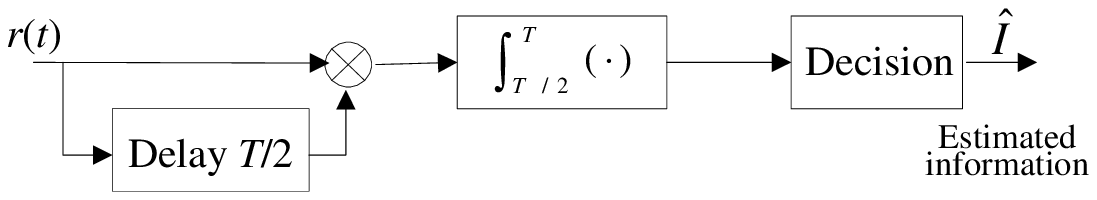}}
\caption{Block diagram of the DCSK communication system: transmitter (a) and receiver (b).}
\label{fig:Fig.1}  
\end{figure}

DCSK modulation employs a chaotic signal as the carrier, with differential shift keying modulators, for transmission. When the chaotic signal is generated by a discrete-time chaotic circuit, the global spreading factor is $2 \beta$ \cite{lau2003chaos}. Fig.~\ref{fig:Fig.1} shows the block diagram of the binary DCSK communication system. As seem from this figure, the binary DCSK modulation unit transmits a reference segment of the chaotic signal in the first half of the symbol duration and, according to the bit information, it repeats or reverses the segment in the last half of the duration to ``$1$'' or ``$0$'' respectively. The $l^{\rm th}$ binary transmitted signal is represented by a sequence of samples of the chaotic signal whose length is $2 \beta$, in which the $j^{\rm th}$ sample is given by
\begin{equation}
\hspace{-3mm}s_j  = \left\{ \begin{array}{rl} \hspace{-2mm} c_j &\mbox{$j = 2 (l - 1) \beta + 1,\cdots,(2 l - 1) \beta$} \\
       \hspace{-2mm}(2b_l-1)c_{j - \beta} &\mbox{$j =(2 l - 1) \beta + 1,\cdots,2 l \beta$}
\end{array} \right.
\label{eq:modu_signal}
\end{equation}
where the $l^{\rm th}$ transmitted symbol is denoted by
\begin{equation}
b_l =  \{0, 1\}
\label{eq:trans_bit}
\end{equation}
Here, ``$0$'' and ``$1$'' appear with the same probability.

The output signal of the multipath fading channel, represented by $r_j(t)$, can be expressed as
\begin{equation}
r_j(t) = \sum_{i=1}^{L} \alpha_i s_{j - \tau_i} + n_j, \quad j = 2 (l - 1) \beta + 1,\cdots,2 l \beta
\label{eq:receiver}
\end{equation}
where $\alpha_i$ and $\tau_i$ are the independent Nakagami-distributed random variable and the time delay of the $i^{\rm th}$ path, respectively, $L$ is the number of fading paths, and $n_j$ is the Gaussian-distributed random noise with the power spectral density of $N_0$. The demodulated signal is given by
\begin{equation}
z = \int_{T/2}^{T} r(t) r(t - T / 2) {\rm d} t
\label{eq:demodu_signal}
\end{equation}
where $T$ is the duration period of a bit.
\subsection{DCSK System Based on Walsh Code} \label{sect:multi_DCSK}
For a multi-access DCSK communication system,
the orthogonal Walsh code sequences are adopted to ensure the orthogonality of the signals among different users \cite{1015003}. However, the interference among different users can be neglected if the Walsh code is synchronized.

\begin{figure}[htbp]
\center
\includegraphics[width=3.5in,height=1.2in]{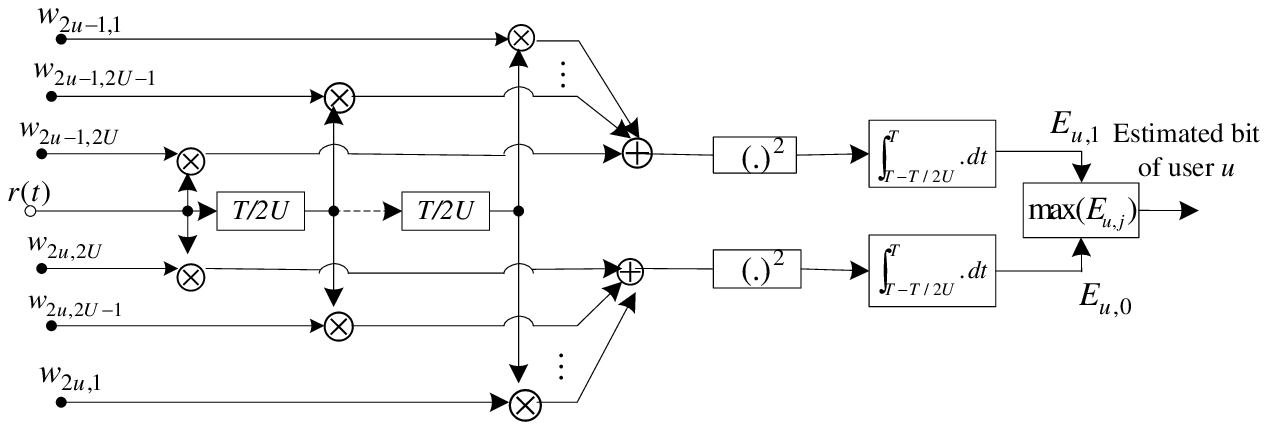}
\vspace{-0.4cm}
\caption{ Block diagram of the GML detector for the $u^{\rm th}$ user.}
\label{fig:Fig.2}
\end{figure}

Consider a system with $U$ users. Let $f$ denote the length of each carrier segment and $2 \beta$ denote the global spreading factor. The $2U$-order Walsh code is used to accommodate the $U$ users. During the modulation, one keeps
 $f = 2 \beta / 2U $ in order to ensure the global spreading factor $2 \beta$ be constant. The $l^{\rm th}$ transmitted signal of the $u^{\rm th}$ user can be expressed by the $2U$-order orthogonal Walsh code, as follows:
\begin{equation}
s_{u, b_l} = \sum_{i=0}^{2U-1} w_{2u-b_l,i}\,c(t-i \frac{T} {2U}), \quad 0 < t < T
\label{eq:multi_signal}
\end{equation}
where $b_l$ is the $l^{\rm th}$ transmitted symbol and $w_k$, ($k=1, 2,\cdots,2U$) is a row vector of the $2U$-order Walsh code.
During the demodulation process, the generalized maximum likelihood (GML) detector \cite{1329074} should be applied instead of the differentially coherent detector. The block diagram of the GML detector for the
$u^{\rm th}$ user is plotted in Fig.~\ref{fig:Fig.2}, and the weighted energy combined of the $l^{\rm th}$ received signal of the $u^{\rm th}$ user is expressed as
\begin{equation}
E_{u,b_l} = \int_{T-T/2U}^{T} \left[ \sum_{i=0}^{2U-1}r (t-i \frac{T} {2U}) w_{2u-b_l,2U-i} \right]^2 {\rm d} t
\label{eq:GML_demodu}
\end{equation}
Here, $r (t-iT/ 2U)$ represents the received signal which has delay $iT/ 2U$.

For example, to accommodate $2$ users, the system should use the $4$-order Walsh code sequences where
\begin{eqnarray}
\mW_{4} =
\left[
\begin{array}{l}
w_1\\
w_2\\
w_3\\
w_4\\
\end{array}
\right] =
\left[
\begin{array}{llll}
+1 & +1 & +1 & +1\\
+1 & -1 & +1 & -1\\
+1 & +1 & -1 & -1\\
+1 & -1 & -1 & +1\\
\end{array}
\right]
\label{eq:Walsh_2}
\end{eqnarray}

Each user selects a couple of vectors to form the transmitted signal. User $1$ can take $w_1$ and $w_2$
for bit ``$1$'' and ''$0$'', and user $2$ can adopt the rest two vectors.
\subsection{Two-user MIMO Relay DCSK-CD System} \label{sect:DCSK-CD_system}
Consider a two-hop network system model with two users (sources), one destination and one relay. Both the relay and destination can employ multiple antennas while each user possesses only one antenna.
The channels constructed by the transmit-receive antenna pairs are independent and subjected to static block frequency-selective fading, meaning that the channel state remains constant during each transmission period.
 In this system, the users do not cooperate with each other in sending messages but only the relay helps them to transmit messages to the destination. To simplify the analysis, it is assumed that a transmission period is divided into broadcast phase and cooperative phase, denoted as
$1^{\rm st}$ time slot and $2^{\rm nd}$ time slot, respectively. In the $1^{\rm st}$ time slot, users broadcast their messages to other terminals (relay and destination),
and then the relay cooperates with two users to send their messages to the destination in the $2^{\rm nd}$ time slot.
The basic system model is presented in Fig.~\ref{fig:Fig.3}, where each user can support only one transmit antenna,
the relay possesses $M_R$ antennas used under ideal condition (error free (EF) at the relay antennas) or
decode-and-forward (DF) protocol\cite{4273702},
and the destination has $M_D$ receive antennas. The transmission principle of the system is shown in Fig.~\ref{fig:Fig.4},
where $X_{RA}(t)$ and $X_{RB}(t)$ represent the information reconstructed by the relay.
In particular, $X_{RA}(t)=X_A(t)$ and $X_{RB}(t)=X_B(t)$ for the EF situation; for the DF case,
 $X_{RA}(t)=X_A(t)$ and $X_{RB}(t)=X_B(t)$ if the relay is able to decode the signal correctly, and $X_{RA}(t)=X_{RB}(t)=0$ otherwise.
The function $F(X_{RA}(t),X_{RB}(t))$ describes
the processing of the reconstructed signals. In the proposed system, $F(X_{RA}(t),X_{RB}(t)) = X_{RA}(t)+X_{RB}(t)$.
Consequently, the received signals at the relay and destination antennas can be expressed as follows:
\begin{figure}[htbp]
\center
\includegraphics[width=3.0in,height=1.3in]{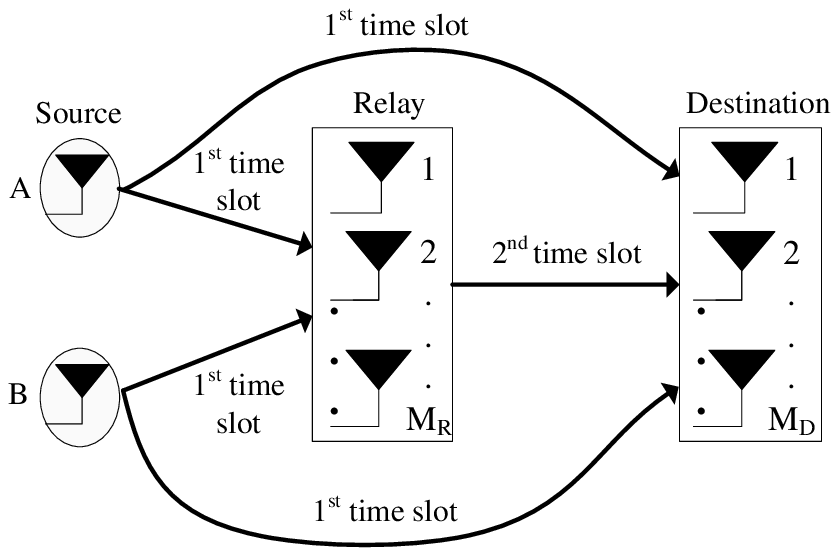}
\vspace{-0.3cm}
\caption{The basic model of the $2$-user MIMO relay DCSK-CD system.}
\label{fig:Fig.3}
\end{figure}
\begin{figure}[htbp]
\center
\includegraphics[width=3.0in,height=1.2in]{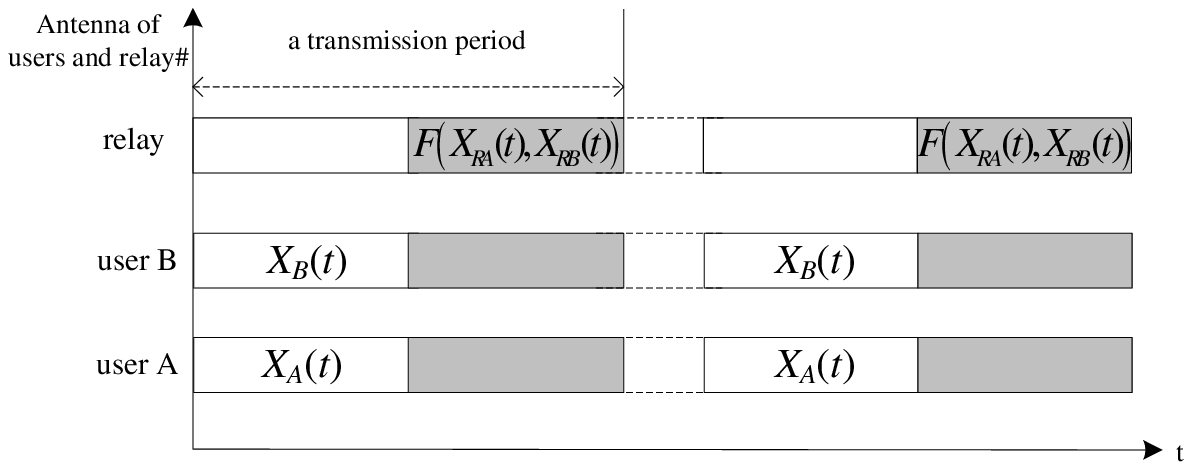}
\vspace{-0.3cm}
\caption{The transmission principle of the proposed system.}
\label{fig:Fig.4}
\end{figure}

\quad $1^{\rm st}$ time slot:
\begin{equation}
\begin{split}
&r_R^1 (t)= H_{AR} \otimes X_A (t) + H_{BR} \otimes X_B (t)  + Z_R^1 (t)\\
&r_D^1 (t)= H_{AD} \otimes X_A (t) + H_{BD} \otimes X_B (t)  + Z_D^1 (t)
\label{eq:DCSK-CD_receiver}
\end{split}
\end{equation}

\quad $2^{\rm nd}$ time slot:
\begin{equation}
\hspace{-9mm}
r_D^2 (t)= H_{RD} \otimes ( X_{RA} (t) + X_{RB} (t) )  + Z_D^2 (t)
\label{eq:DCSK-CD_receiver}
\end{equation}
Here, the superscripts ``$1$'' and ``$2$'' are used to denote the $1^{\rm st}$ time slot and the $2^{\rm nd}$ time slot, $\otimes$ represents the convolution operator, $r_D^1 (t)$ and $r_D^2 (t)$ are the total received signals at destination for users A and B in the $1^{\rm st}$ time slot and the $2^{\rm nd}$ time slot, respectively, $r_R^1 (t)$ is the total received signals at relay for two users in the $1^{\rm st}$ time slot. $X_A(t)$ and $X_B (t)$ is the signal transmitted by user A and B, $Z_R^i (t)$ and $Z_D^i (t)$ ($i=1,2$) are white Gaussian noise random variables with zero mean and the power spectral density $N_0$. The channel multipath impulse response $H_{mn}$ are modeled by a linear time-invariant process, i.e., $H_{mn}= \sum_{i=1}^{L} \alpha_i \delta (t - \tau_i )$, where $\alpha_i$ and $\tau_i$  denote the attenuation and time delay of the $i^{\rm th}$ path, respectively, and $L$ represents the number of the multipath taps.

The channel coefficients $\mid \alpha_i\mid$ follows a Nakagami distribution, with the probability density function (PDF) given as
\begin{equation}
f_{\mid \alpha_i\mid} (\alpha)= \frac{2} {\Gamma (m_i)} \left( \frac{m_i} {\Omega_i} \right)^{m_i} \alpha^{2 m_i - 1}
\exp \left( - \frac{m_i} {\Omega_i} \alpha^2 \right)
\label{eq:Nakagami_PDF}
\end{equation}
where $m_i$ is the fading parameter of the $i^{\rm th}$-path, $\Omega_i$ equals $E(\alpha_i^2)$, and $\Gamma(\cdot)$ denotes the Gamma function. To simplify the analysis, it is assumed that each fading channel has a uniform scale parameter, meaning that $\Omega_i / m_i$ is constant for all the paths. Meanwhile, the receiver knows all multipath parameters, thus $\sum_{i=1}^{L} \Omega_i= \sum_{i=1}^{L} E(\alpha_i^2)=1$.

\begin{figure}[tbp]
\center
\includegraphics[width=3.0in,height=1.25in]{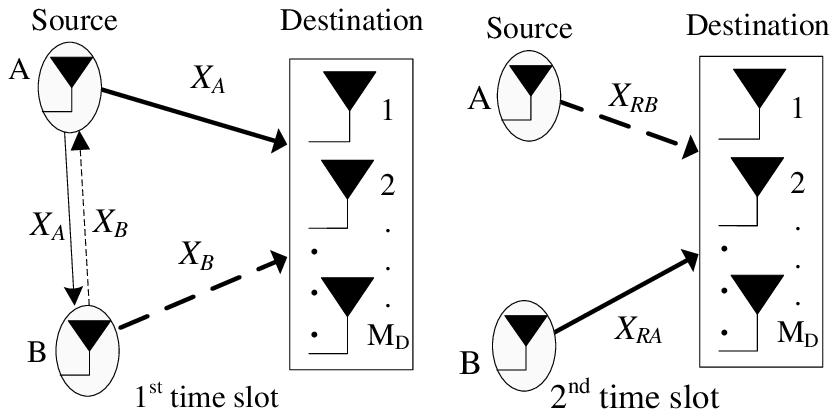}
\vspace{-0.3cm}
\caption{The basic model of the $2$-user DCSK-CC system.}
\label{fig:Fig.5}
\end{figure}

As two conventional DCSK systems, the DCSK-NC system does not use any cooperative technique; for the DCSK-CC system \cite{5629387}, which involves no relays, the users do cooperate with each other to transmit messages.
Fig.~\ref{fig:Fig.5} shows the basic model of the $2$-user DCSK-CC system, where the parameters are shown in the first paragraph of this subsection. Referring to this figure, one can observe that the DCSK-CD system
(see Fig.~\ref{fig:Fig.3}) is quite different from the DCSK-CC system. In the DCSK-CC system, the users broadcast their messages to other terminals (the other user and destination) in the $1^{\rm st}$ time slot.
Then, these two users cooperate with each other to send their messages to the destination through two different channels in the $2^{\rm nd}$ time slot.

Comparing with the DCSK-CC system, a pair of transmit-receive antennas can be used to construct an independent fading channel in the proposed DCSK-CD system
 without increasing much complexity.
 Thus, the anti-fading capacity can be significantly enhanced. As a result, a considerable performance gain is obtained by
 incorporating the MIMO relay technology into the DCSK system.

\section{Performance Analysis} \label{sect:Performance}
In this section, the exact BER expression of DCSK-NC system \cite{1362943} is re-derived over Nakagami fading channels in order to obtain the bit error rate (BER) formula about the multi-access MIMO relay DCSK-CD system. Furthermore, a method is presented
 in closed-form for approximating the BER expressions of the DCSK-NC and the proposed system. In the following analysis, we assume that the energy per bit is constant, denoted by $E_b$, the Walsh code used in the system has been perfectly synchronized,
 and the time delay of each path $\tau_i$ is much shorter than the bit duration
 ($\tau_i \ll 2 \beta T_s$, $T_s$ is the sampling period of the chaotic signal),
 so that the inter-symbol interference (ISI) in the transmission can be neglected.

\subsection{Exact BER of Conventional DCSK-NC System} \label{sect:exactconv_BER}
First, the BER expression of the DCSK-NC system is re-derived over Nakagami fading channels.

The signal-to-noise ratio (SNR) per bit of the $\gamma_b$ at the receiver can be expressed as
\begin{equation}
\gamma_b = \sum_{i=1}^{L}  \gamma_i = \frac{E_b} {N_0} \sum_{i=1}^{L} \alpha_i^2
\label{eq:conv-rece_SNR}
\end{equation}
where the bit energy $E_b$ is defined by $E_b=2\beta E[(c_j)^2]$, $N_0$ is the power specral density of noise.
$L$ is the number of independent fading paths, $\gamma_i$ is the instantaneous SNR of the $i^{\rm th}$ path.
and $\alpha_i$ is the Nakagami-distributed varaible with parameter $m_i$.

It is well known that the square of a Nakagami-distributed random variable follows a gamma distribution.
The PDF of a gamma-distributed variable $X$ with parameters $a>0, b>0$,
denoted as $X\sim G(a, b)$, is given by
\begin{equation}
f (x) = \frac{x^{a-1} e^{-x/b}} {b^a \Gamma(a)}, \quad x>0
\label{eq:Gamma_PDF}
\end{equation}
As $\alpha_i$ is the Nakagami-distributed variable with parameter $m_i$,
we have $\gamma_i = (E_b/N_0) \alpha_i^2 \sim G \left(m_i, (E_b/N_0) \Omega_i/m_i \right)$.
The MGF of the variable $X\sim G(a, b)$ is given by  \cite{Simon2005Rayl}
\begin{equation}
M(s) = (1 - s b)^{-a}
\label{eq:Gamma-MGF}
\end{equation}
Since $\gamma_1, \cdots ,\gamma_L$ are statistically independent and $\Omega_i / m_i$ is constant,
we can get the MGF of $\gamma_b = \sum_{i=1}^{L}\gamma_i$ through \cite{Simon2005Rayl}
\begin{equation}
M_{\gamma_b}(s) = \displaystyle \prod_{i=1}^{L}(1 - s \frac{ (E_b/N_0) \Omega_i} {m_i})^{-m_i}
\label{eq:Y-MGF}
\end{equation}
Thus the distribution of $\gamma_b$ can be obtained using the inverse MGF transform of \eqref{eq:Y-MGF}, resulting in
\begin{equation}
\gamma_b \sim G \left(\sum_{i=1}^{L} m_i, \frac{ \displaystyle E_b \sum_{i=1}^{L} \Omega_i}
{N_0 \displaystyle \sum_{i=1}^{L} m_i} \right)
= G \left(mL, \frac{E_b/N_0} {mL} \right)
\end{equation}
where $m_i (m_1 = m_2, \cdots ,= m_L = m)$ is the Nakagami fading parameter of the $i^{\rm th}$ path which controls the depth of the fading envelope.
Hence, the PDF of $\gamma_b$  can be written as
\begin{equation}
f (\gamma_b) = \frac{\gamma_b^{mL-1} e^{-\gamma_b / \left(\frac{E_b/N_0}
 {mL} \right)}} {\left(\frac{E_b/N_0} {mL} \right)^{mL} \Gamma (mL)}
\label{eq:convrece_PDF}
\end{equation}
The conditional BER of the DCSK-NC system as a function of the received SNR is given by \cite{1362943}
\begin{equation}
\begin{array}{rl}
{\rm BER} (\gamma_b) =& \hspace{-2mm} \displaystyle \frac{1} {2}\, \text{erfc} \left( \left[\frac{4} {\gamma_b}
\left(1 + \frac{f} {2 \gamma_b}\right)^{- \frac{1} {2}} \right] \right)
\vspace{1mm}\\
=& \hspace{-2mm} \displaystyle Q \left( \sqrt{\frac{\gamma_b^2} {2 \gamma_b + f}} \right)
\end{array}
\label{eq:conv-CBER}
\end{equation}
where $f$ is the length of each carrier segment, $\text{erfc}(\cdot)$ is
defined as
\begin{equation}
\text{erfc} (x) = \frac{2} {\sqrt{\pi}} \int_{x}^{\infty} e^{-\lambda^2} {\rm d}\lambda
\label{eq:erfc_fuction}
\end{equation}
and $Q(\cdot)$ is the Gaussian Q-fuction.
Combining \eqref{eq:convrece_PDF} and \eqref{eq:conv-CBER}, the BER of the DCSK-NC system can be obtained by averaging the conditional BER, as follows:
\begin{equation}
{\rm BER} =  \int_{0}^{\infty} {\rm BER}(\gamma_b) f(\gamma_b) {\rm d} \gamma_b
\label{eq:BER-expresion}
\end{equation}
This formula will be used for deriving the exact BER expressions of the MIMO relay
DCSK-CD system with different relay protocols in the next section.
\subsection{Exact BER of Multi-access MIMO Relay DCSK-CD System} \label{sect:DCSKCD-BER}
In the multi-access MIMO relay DCSK-CD system, assume that the fading coefficient $\alpha_i$ of each path is independent and identically-distributed (i. i. d.). Consider a system includes $n$ users with only one antenna, one receiver with $M_D$ antennas and one relay with $M_R$ antennas. To simplify the analysis, also assume that the transmitted messages arrived at each receive antenna possess identical energy and all independent channels are subjected to the same channel condition. Moreover, let $2\beta$ denote the global spreading factor, and the $d_{SD}$, $d_{SR}$, $d_{RD}$ denote the distance of source to destination (S-D), source to relay (S-R) and relay to destination (R-D) links, respectively, thus $d_{SD}:d_{SR}:d_{RD}$
represents the geometric positions of the users.
The path loss of the S-D link is given by $ PL(S,D)=1/d_{SD}^2$.

For easily deriving the the exact BER of the multi-access MIMO relay DCSK-CD system with different protocols,
we should introduce the following two theorems.
\begin{theorem}
\emph{Given $N$ independent gamma random variables $X_1, X_2,\cdots,X_N$, where $X_k \sim G (a_k, b)\,(k=1,\cdots,N)$. The sum of these variables
$X=\sum_{k=1}^{N} X_k$ also follows a gamma distribution, denoted as $X \sim G (\sum_{k=1}^{N} a_k, b)$.}
\label{theo:theorem1}
\end{theorem}
\begin{theorem}
\emph{Assume that $N$ independent gamma random variables $X_1, X_2,\cdots,X_N$ are
distributed as $X_k \sim G (a_k, b_k)\,(k=1,\cdots,N)$ where $b_1 \neq b_2,\cdots,\neq b_N$.
The PDF of the sum of these variables $X = \sum_{k=1}^{N} X_k$ is expressed as \cite{Moschopoulos1985}}
\begin{equation}
f(x)=  C \sum_{i=0}^{\infty}\left( \eta_i x^{\rho + i - 1} e^{-x/b_0}/ \left[ \Gamma (\rho + i) b_0^{\rho + 1} \right] \right)
\label{eq:indepsum-Gamma-PDF}
\end{equation}
\emph{which is subjected to}
\begin{equation}
\left\{
\begin{aligned}
& C = \prod_{k=1}^{N} (b_0 / b_k )^{a_k}\\
&\eta _{i + 1} = \frac{1} {i + 1} \sum_{t=1}^{i + 1} t z_t \eta_{i+1-t} &\mbox{$i = 0,1,2,\cdots$}\\
& z_j = \sum_{k=1}^{N} a_k ( 1 - b_0 / b_k)^j/j &\mbox{$j =1,2,\cdots$}\\
& \rho = \sum_{k=1}^{N} a_k > 0\\
& b_0 = \min_k (b_k)
\end{aligned}
\right.
\label{eq:indepsum-Gamma-parameters}
\end{equation}
\emph{where $\eta_0 = 1$.}
\label{theo:theorem2}
\end{theorem}
\subsubsection{BER with EF protocol for the relay (ideal condition)} \label{sect:EF-DCSKCD-BER}

Assume that the relay can receive messages from the users without errors, namely, the BER of S-R link is zero. Obviously, there are $M_D$, $M_R$ and $M_R M_D$ independent Nakagami fading channels for S-D link, S-R link and R-D link. The conditional BER and total BER expressions of the proposed system are expressed by \eqref{eq:conv-CBER} and \eqref{eq:BER-expresion}. Thus, what needs to do is to deduce the PDF of the total received SNR. Let the received SNRs for the relay and the destination in the $1^{\rm st}$ time slot be denoted by $\gamma_{SR}$ and $\gamma_{SD}$, and similarly the received SNR for the destination in the $2^{\rm nd}$ time slot be denoted by $\gamma_{RD}$.

In the $1^{\rm st}$ time slot,
 namely the broadcast phase, based on the equivalent transmitted SNR of this channel is $(E_b / 2M_D)/N_0$, one can easily obtain the distribution of the received SNR for the $k^{\rm th}$ source-destination antenna pair, as follows:
\begin{equation*}
\gamma_{SD-k} \sim G \left(mL,  (E_b / N_0)/(2 M_D d_{SD}^2 m n L)\right)
\end{equation*}
Thus, the distribution of $\gamma_{SD}$
can be deduced using Theorem~\ref{theo:theorem1},
 shown as
\begin{equation}
\begin{array}{rl}
\hspace{-4mm}\gamma_{SD} = \displaystyle \sum_{k=1}^{M_D} \gamma_{SD-k}
 \sim & \displaystyle \hspace{-1mm}  G \left( M_D m L, \frac{E_b / N_0} {2 M_D d_{SD}^2 m n L} \right)
\vspace{1mm}\\
 =& \hspace{-1mm} G (a_1, b_1)
 \end{array}
\label{eq:SD-receSNR}
\end{equation}
where $m$ is the Nakagami fading parameter of each path, $n$ is the number of users. In the following, $a_1$ and $b_1$ are used as short-hand notations for $M_D m L$ and
$(E_b / N_0) / (2 M_D d_{SD}^2 m n L)$, respectively.

Similarly, in the $2^{\rm nd}$ time slot,
 one can get the distribution of received SNR for the $k^{\rm th}$ antenna pair for R-D link,
namely $\gamma_{RD-k} \sim G \left(mL, (E_b / N_0)/( 2 M_R M_D d_{RD}^2 m n L)\right)$.
 Thus, $\gamma_{RD}$ is given by
\begin{equation}
\begin{array}{rl}
\hspace{-2mm} \gamma_{RD} = \hspace{-2mm} \displaystyle \sum_{k=1}^{M_R M_D} \hspace{-2mm} \gamma_{RD-k}
\sim & \displaystyle \hspace{-2mm} G \left( M_R M_D m L, \frac{E_b / N_0} {2 M_R M_D d_{RD}^2 m n L} \right) \vspace{1mm}\\
=& \hspace{-2mm} G (a_2, b_2)
\end{array}
\label{eq:RD-receSNR}
\end{equation}
Here, likewise, $a_2$ and $b_2$ are short-hand notations for $M_R M_D m L$ and
$(E_b / N_0) / (2 M_R M_D d_{RD}^2 m n L)$.

Finally, using the equal-gain combining method, the total received SNR for destination with two
signals from the relay and one source $\gamma_D$ can be formulated as
\begin{equation}
\gamma_D = (\gamma_{SD} + \gamma_{RD})/\sqrt{2}
\label{eq:SDRD-receSNR}
\end{equation}
where dividing by $\sqrt{2}$ (normalized factor) is used
to keep the total received energy for the destination from two different terminals in the
proposed system be consistent with the ones of the DCSK-NC system.

Combining \eqref{eq:SD-receSNR}-\eqref{eq:SDRD-receSNR} with Theorems \ref{theo:theorem1} and \ref{theo:theorem2},
we can write the PDF of $\gamma_D$ as
\begin{equation}
\hspace{-0.5mm} f (\gamma_D)  = \left\{\displaystyle \begin{array}{rl}
\frac{\displaystyle \gamma_D^{a_1 + a_2 - 1} e^{- \gamma_D / (b_1/ \sqrt{2})}} {\displaystyle (b_1/ \sqrt{2})^{a_1+a_2} \Gamma (a_1+a_2)} &\mbox{ if $b_1 = b_2$}\vspace{1mm}\\
C {\displaystyle \sum_{i=0}^{\infty}} \left( \frac{\displaystyle \eta_i \gamma_D^{\rho + i - 1} e^{- \gamma_D / b_0}} {\displaystyle  \Gamma (\rho + i) b_0^{\rho + 1}} \right) &\mbox{ if $b_1 \neq b_2$}
\end{array} \right.
\label{eq:SDRD-PDF}
\end{equation}
where all parameters are as shown in \eqref{eq:indepsum-Gamma-parameters}-\eqref{eq:RD-receSNR}.

Substituting \eqref{eq:conv-CBER} and \eqref{eq:SDRD-PDF} into \eqref{eq:BER-expresion}, one can get the exact BER formula of the proposed system
under the ideal condition as \eqref{eq:EF-BER},
\begin{figure*}[htbp]
\begin{eqnarray}
{\rm BER}_{\rm EF} \hspace{-1.5mm} &=& \hspace{-1.5mm} {\rm BER}_{D} = \int_{0}^{\infty} {\rm BER}(\gamma_D) f(\gamma_D) {\rm d} \gamma_D \cr
&=& \hspace{-1.5mm} \left\{\displaystyle \begin{array}{rl}
\displaystyle \int_{0}^{\infty} \frac{ \gamma_D^{a_1 + a_2 - 1} e^{- \gamma_D / (b_1/ \sqrt{2})}}
{(b_1/ \sqrt{2})^{a_1+a_2} \Gamma (a_1+a_2)}
Q \left( \sqrt{\frac{\gamma_D^2} {2 \gamma_D + f}} \right) {\rm d} \gamma_D &\mbox{ if $b_1 = b_2$}\vspace{1mm}\\
\displaystyle \int_{0}^{\infty} \left( C {\displaystyle \sum_{i=0}^{\infty}} \left( \frac{\displaystyle \eta_i \gamma_D^{\rho + i - 1} e^{- \gamma_D / b_0}} {\displaystyle  \Gamma (\rho + i) b_0^{\rho + 1}} \right) \right)
Q \left( \sqrt{\frac{\gamma_D^2} {2 \gamma_D + f}} \right) {\rm d} \gamma_D &\mbox{ if $b_1 \neq b_2$}
\end{array} \right.\cr
&=& \hspace{-1.5mm} \left\{\displaystyle \begin{array}{rl}
\displaystyle \frac{1} {(b_1/ \sqrt{2})^{a_1+a_2} \Gamma (a_1+a_2)}
\int_{0}^{\infty} \gamma_D^{a_1 + a_2 - 1} e^{- \gamma_D / (b_1/ \sqrt{2})}
Q \left( \sqrt{\frac{\gamma_D^2} {2 \gamma_D + f}} \right) {\rm d} \gamma_D &\mbox{ if $b_1 = b_2$}\vspace{1mm}\\
\displaystyle C \int_{0}^{\infty} \left( {\displaystyle \sum_{i=0}^{\infty}} \left( \frac{\displaystyle \eta_i \gamma_D^{\rho + i - 1} e^{- \gamma_D / b_0}} {\displaystyle  \Gamma (\rho + i) b_0^{\rho + 1}} \right) \right)
Q \left( \sqrt{\frac{\gamma_D^2} {2 \gamma_D + f}} \right) {\rm d} \gamma_D &\mbox{ if $b_1 \neq b_2$}
\end{array} \right.
\label{eq:EF-BER}
\end{eqnarray}
\end{figure*}
where $f$ is the length of each carrier segment.

\subsubsection{BER with DF protocol for the relay} \label{sect:DF-DCSKCD-BER}
As one of the most popular relaying techniques employed in cooperative diversity systems, DF protocol performs such that if the relay can decode the received signal correctly, then it forwards the decoded message to the destination
in the $2^{\rm nd}$ time slot, otherwise, it remains idle.
 The BER of the DF case is based on the one of the EF protocol, described as
\begin{equation}
{\rm BER}_{\rm DF}  = {\rm BER}_{SR} \cdot {\rm BER}_{SD} + (1 - {\rm BER}_{SR}) \cdot {\rm BER}_{D}
\label{eq:DF-BER}
\end{equation}
where ${\rm BER}_{SR}$, ${\rm BER}_{SD}$ and ${\rm BER}_{D}$ denote the BER of the relay receiver with one signal from one source, the BER of the destination receiver
with one signal from one source, and the total BER of the destination receiver with two signals from one source and the relay, respectively.
The distribution of the received SNR for the relay is
\begin{equation}
\begin{array}{rl}
\hspace{-4mm}\gamma_{SR} =\displaystyle \sum_{k=1}^{M_R} \gamma_{SR-k}
\sim & \displaystyle \hspace{-2mm} G \left( M_R m L, \frac{E_b / N_0} {2 M_R d_{SR}^2 m n L} \right)
\vspace{1mm}\\
 =& \hspace{-2mm} G (a_3, b_3)
 \end{array}
\label{eq:SR-receSNR}
\end{equation}
where $\gamma_{SR-k}$ is the received SNR for the $k^{\rm th}$ S-R channel, expressed as $\gamma_{SR-k} \sim G \left(m L, (E_b / N_0) / (2 M_R d_{SR}^2 m n L) \right)$,
 and and $a_3$ and $b_3$ are short-hand notations for
$M_R m L$ and $(E_b / N_0) / (2 M_R d_{SR}^2 m n L)$.

The distributions of $\gamma_{SD}$ and $\gamma_{D}$ here are the same as the ones under the ideal condition, with expressions shown in \eqref{eq:SD-receSNR} and \eqref{eq:SDRD-PDF} respectively. Then we can obtain the PDFs of these three variables, denoted by $f (\gamma_{SR})$, $f (\gamma_{SD})$, and $f (\gamma_{D})$. The conditional BERs for these variables are identical, as shown in \eqref{eq:conv-CBER}.
Accordingly, the expression of ${\rm BER}_{D}$ is shown in \eqref{eq:EF-BER},
while ${\rm BER}_{SR}$ and ${\rm BER}_{SD}$ can be calculated by \eqref{eq:BER-expresion}, thus we have
\begin{equation}
\begin{split}
\hspace{-6mm} &{\rm BER}_{SR} = \hspace{-0mm} \displaystyle \int_{0}^{\infty} \frac{ \gamma_{SR}^{a_3-1} e^{- \gamma_{SR} / b_3}}
{b_3^{a_3} \Gamma (a_3)}
Q \left( \sqrt{\frac{\gamma_{SR}^2} {2 \gamma_{SR} + f}} \right) {\rm d} \gamma_{SR}\vspace{0.5mm}\\
&= \hspace{-0mm} \displaystyle \frac{1} {b_3^{a_3} \Gamma (a_3)} \int_{0}^{\infty} \gamma_{SR}^{a_3-1} e^{- \gamma_{SR} / b_3}
\displaystyle Q \left( \sqrt{\frac{\gamma_{SR}^2} {2 \gamma_{SR} + f}} \right) {\rm d} \gamma_{SR}
\label{eq:SR-BER}
\end{split}
\end{equation}
\begin{equation}
\begin{array}{rl}
\hspace{-5mm} {\rm BER}_{SD}
=& \hspace{-3mm} \displaystyle \frac{1} {b_1^{a_1} \Gamma (a_1)} \int_{0}^{\infty} \gamma_{SD}^{a_1-1} e^{- \gamma_{SD} / b_1} \vspace{1mm}
\\& \displaystyle \hspace{-3mm} \times Q \left( \sqrt{\frac{\gamma_{SD}^2} {2 \gamma_{SD} + f}} \right) {\rm d} \gamma_{SD}
\end{array}
\label{eq:SD-BER}
\end{equation}

Finally, substituting \eqref{eq:EF-BER}, \eqref{eq:SR-BER} and \eqref{eq:SD-BER} into \eqref{eq:DF-BER}, the BER of the proposed system with DF relay protocol can be formulated.
\subsection{Approximate Results of the DCSK-NC System and the Proposed System} \label{sect:DCSK-APPBER}
Although the exact average BER can be computed through numerical evaluation of the double integral in \eqref{eq:BER-expresion} with different PDFs of received SNRs. It is very difficult to calculate the integrations of \eqref{eq:BER-expresion} and \eqref{eq:EF-BER}
arising from the nonlinear behavior of the function. To resolve this problem, an approximated method based on the MGF \cite{Simon2005Rayl} is proposed to simplify the expression of the BER into a closed-form.
\subsubsection{DCSK-NC system} \label{sect:DCSK-NC-APP}
To begin with, define $w$ as a function of $\gamma_b$, as
\begin{equation}
w= \frac{\gamma_b^2} {2 \gamma_b + f}
 \label{eq:w-expression}
 \end{equation}
Substituting \eqref{eq:w-expression} into \eqref{eq:BER-expresion} gives the BER expression via averaging the conditional BER of $w$
\begin{equation}
\begin{array}{rl}
{\rm BER} =& \displaystyle \int_{0}^{\infty} Q (\sqrt{w}) f_w (w) {\rm d} w
\\
=& \displaystyle \frac{1} {\pi} \int_{0}^{\pi/2} \int_{0}^{\infty} f_w (w) e^{- \frac{w} {2 \text{sin}^2 \theta}} {\rm d} w {\rm d} \theta
\end{array}
\label{eq:BER-Simp}
\end{equation}
where the Gaussian $Q$-function may be represented as $Q (\sqrt{w}) =  (1/\pi)
\int_{0}^{\pi/2} e^{- \frac{w} {2 \text{sin}^2 \theta}}{\rm d} \theta$.

Since the MGF of $w$ is the Laplace transform of $f_w(w)$ with the exponent reversed in sign, defined as $M_w(s) = \int_{0}^{\infty} e^{sw} f_w(w){\rm d} w$,
putting it into \eqref{eq:BER-Simp} yields
\begin{equation}
{\rm BER}= \frac{1} {\pi} \int_{0}^{\pi/2} M_w \left(- \frac{1} {2 \text{sin}^2 \theta} \right) {\rm d} \theta
\label{eq:BER-MGF-expression}
\end{equation}
Assume that $\gamma_b \sim G(a,b)$. Then, the variable $w$ can be approximately described by a Gamma distribution \cite{4336107}, denoted as $G(a_w,b_w)$, with parameters given by
\begin{equation}
\left\{
\begin{aligned}
& a_w =& \hspace{-2mm} a \left( \frac{\bar{\gamma} + f/2} {\bar{\gamma} + f} \right)^2\\
& b_w =& \hspace{-2mm} \frac{ b \bar{\gamma} (\bar{\gamma} + f)^2} {2 (\bar{\gamma} + f/2)^3}
\end{aligned}
\right.
\label{w-Gamma-parameters}
\end{equation}
where $\bar{\gamma} = \sum_{i=1}^{L} \bar{\gamma}_i = (E_b/N_0) \sum_{i=1}^{L} E(\alpha_i^2) = E_b/N_0$ and $f$ is the length of each carrier segment.

The MGF of $w \sim G(a_w,b_w)$ can be formulated using \eqref{eq:Gamma-MGF}
\begin{equation}
M_w(s) = (1 - s b_w)^{-a_w}
\label{eq:w-MGF}
\end{equation}
Substituting \eqref{eq:w-MGF} into \eqref{eq:BER-MGF-expression} gives the following expression:
\begin{equation}
{\rm BER} = \frac{1} {\pi} \int_{0}^{\pi/2} \left( 1 + \frac{b_w} {2 \text{sin}^2 \theta} \right)^{-a_w} {\rm d} \theta
\label{eq:convAPP_inte}
\end{equation}
which can be written in closed-form by using the definite integral derived in \cite{Simon2005Rayl}. Hence, the approximate result of the BER formula of the DCSK-NC system is
\begin{equation}
\begin{array}{rl}
{\rm BER} \approx&\hspace{-2mm} \displaystyle \frac{1} {2 \sqrt{2 \pi}} \cdot \frac{\sqrt{b_w} \,
\Gamma \left(a_w + \frac{1} {2} \right)} {\left(1 + \frac{b_w} {2}\right)^{a_w +
\frac{1} {2}} \,\Gamma (a_w + 1)} \vspace{1mm}\\&
\displaystyle \times _2 F_1 \left(1, a_w + \frac{1} {2}; a_w + 1; \frac{2} {2 + b_w}\right)
\end{array}
\label{eq:convBER-APP}
\end{equation}
where $_2 F_1 (\cdot, \cdot; \cdot; \cdot)$ is the Gaussian hypergeometric function shown in \cite{abramowitz1964handbook}. It should be noted that the approximate result will become more accurate as the number of paths increases.

\subsubsection{MIMO DCSK-CD system} \label{sect:DCSKCD-APPBER}
We only consider the approximate BER expression with EF protocol
 over the Nakagami fading channel, which is the foundation to derive the one with the DF case.
It is known from \eqref{eq:SDRD-PDF} that $\gamma_D = (\gamma_{SD} + \gamma_{RD})/\sqrt{2}$ is a gamma-distributed variable if $b_1=b_2$, denoted as $\gamma_D \sim G (a_1 + a_2, b_1/\sqrt{2})$. Combining the equation $b_1=b_2$ with \eqref{eq:SD-receSNR} and \eqref{eq:RD-receSNR}, we conclude that $\gamma_D$ follows gamma distribution provided that $d_{SD}/d_{RD} = \sqrt{M_R}$.
In this case, one can also use \eqref{eq:convBER-APP} to estimate the BER of such a system with different parameters $(a_{\rm ef},b_{\rm ef})$ where
\begin{equation}
\left\{
\begin{aligned}
& a_{\rm ef} =&\hspace{-2mm} (a_1 + a_2) \left( \frac{\bar{\gamma}_D + f/2} {\bar{\gamma}_D + f} \right)^2\\
& b_{\rm ef} =&\hspace{-2mm}\frac{ b_1 \bar{\gamma}_D (\bar{\gamma}_D + f)^2} {2 \sqrt{2} (\bar{\gamma}_D + f/2)^3}
\end{aligned}
\right.
\label{EF-w-parameters}
\end{equation}
Here, $\bar{\gamma}_D = \sum_{i=1}^{L} \bar{\gamma}_{iD} = E (\gamma_D)$.
\\Hence, the approximate BER expression is
\begin{equation}
\begin{array}{rl}
{\rm BER}_{\rm EF} \approx&\hspace{-2mm} \displaystyle \frac{1} {2 \sqrt{2 \pi}} \cdot \frac{\sqrt{b_{\rm ef}} \,
\Gamma \left(a_{\rm ef} + \frac{1} {2} \right)} {\left(1 + \frac{b_{\rm ef}} {2}\right)^{a_{\rm ef} +
\frac{1} {2}} \,\Gamma (a_{\rm ef} + 1)} \vspace{1mm}\\&
\displaystyle \times _2 F_1 \left(1, a_{\rm ef} + \frac{1} {2}; a_{\rm ef} + 1; \frac{2} {2 + b_{\rm ef}} \right)
\end{array}
\label{eq:DCSKCD-BER-APP}
\end{equation}

The approximate result of the DCSK-CD system with DF protocol can be obtained similarly. However,
in the case of $b_1 \neq b_2$,
it is not possible to derive the MGF for $\gamma_D$ since it does not follow gamma or any other regular distributions.
Therefore, the close-form BER expression is unavailable today.

\section{Simulation Results} \label{sect:SIMU}
In this section, some numerical calculation, approximation and simulation results of BER are presented. Since the performance of the multi-access system with more than three users is the same as that with two users when the energy of each user is constant, we only consider the systems with two users here.

It is assumed that all channels have the same Gaussian-distributed random noise and the same path-loss model. The transmission energy per bit is kept constant and the energy of each path is allocated uniformly. Unless otherwise stated, the simulations are performed in $2$-user systems over Nakagami fading channels, with parameters $m=1$ and $L=2$.
The delay vector $\tau=(\tau_1,\tau_2)=(0, T_s)$. The global spreading factor of DCSK is much higher than the path delay $\tau_i$,
namely $2\beta T_s \gg \tau_i$.
In the experiments, $(M_R, M_D)$ is used to represent the number of antenna pairs at the relay and the destination, respectively.
\begin{figure}[tbp]
\centering
\subfigure[]{ \label{fig:subfig:a} 
\includegraphics[width=3.0in,height=2.5in]{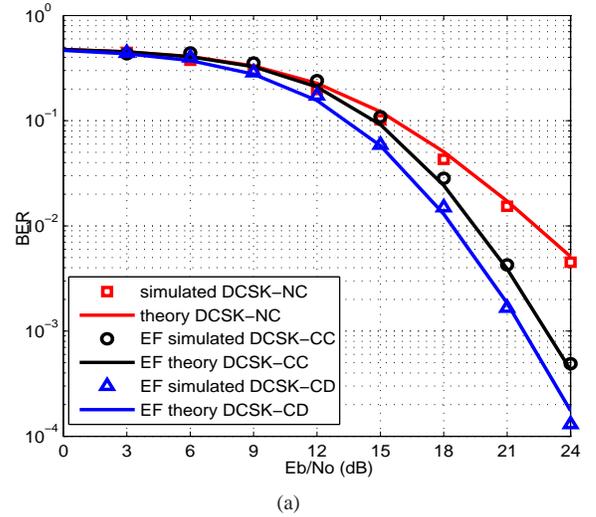}}
\subfigure[]{ \label{fig:subfig:b} 
\includegraphics[width=3.0in,height=2.5in]{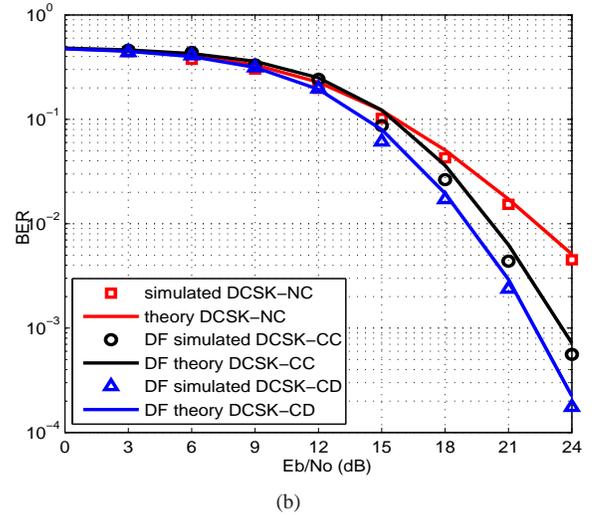}}
\caption{BER curves of three different systems with the EF protocol (a) and the DF protocol (b).}
\label{fig:Fig.6}  
\end{figure}
\subsection{Performance Comparison among the MIMO Relay DCSK-CD, DCSK-CC and DCSK-NC Systems}
\label{subsect:Performance-3methods}
Fig.~\ref{fig:Fig.6} plots the BER curves of the DCSK-NC, the conventional DCSK-CC and
the MIMO relay DCSK-CD systems with EF (the ideal condition) protocol and DF protocol over Nakagami fading channels,
where $d_{SD}: d_{SR}: d_{RD}$ and $(M_R, M_D)$ are set to $1: 1: 1$ and $(1, 1)$, respectively.
The global spreading factor considered here is $2 \beta = 128\,(f=32)$.
Referring to this figure, the BER performance of the proposed DCSK-CD and the DCSK-CC systems is remarkably
better than the DCSK-NC system at a BER of $4\times10^{-3}$. It also shows that the performance of
the DCSK-CD system outperforms the DCSK-CC system by about $1$~dB for BER $= 4\times10^{-3}$ under these two conditions.
Furthermore, we observe that the numerical BER and the simulation results are highly consistent. More performance
gain of the proposed system can be expected as $(M_R, M_D)$ becomes larger.

\begin{figure}[tbp]
\center
\includegraphics[width=3in,height=2.5in]{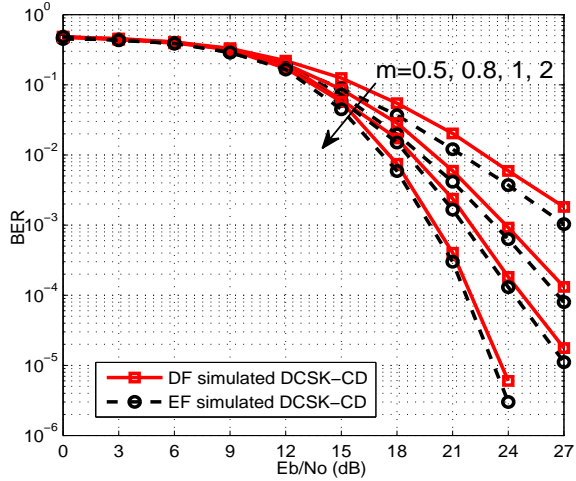}
\vspace{-0.3cm}
\caption{BER curves of the MIMO relay DCSK-CD systems with EF and DF protocols for different fading depths.
}
\label{fig:Fig.7}
\end{figure}

Further, Fig.~\ref{fig:Fig.7} plots the BER curves of the proposed DCSK-CD system with EF and DF protocols for different fading depths.
The parameters used are the same as those in Fig.~\ref{fig:Fig.6} except that the fading parameter $m$ now varies ($m=0.5, 0.8, 1, 2$).
Referring to this figure, the performance gap (distance between the two curves for a given BER) between the EF and DF protocols
becomes larger as the fading depth increases ($m$ decreases), for example, the EF protocol has a performance gain about $1.5$~dB over the DF case at a BER of
$2\times10^{-3}$  when $m=0.5$. However, we also observe that the performance of the proposed system with DF protocol very approaches that of the EF case
when the value of $m$ is over one because the relay can successfully decode most of the transmitted bits.
\begin{figure}[tbp]
\center
\includegraphics[width=3in,height=2.5in]{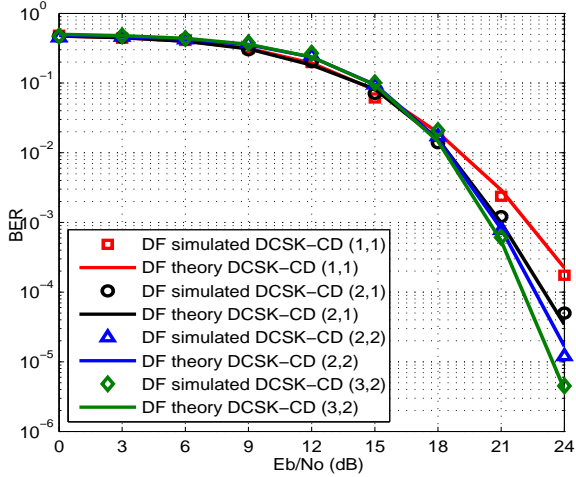}
\vspace{-0.3cm}
\caption{BER curves of the MIMO relay DCSK-CD systems with different antenna pairs.
}
\label{fig:Fig.8}
\end{figure}

\subsection{Performance of Two-user MIMO Relay DCSK-CD System}
\label{subsect:Performance-DCSKCD}
\begin{figure}[tbp]
\center
\includegraphics[width=3in,height=2.5in]{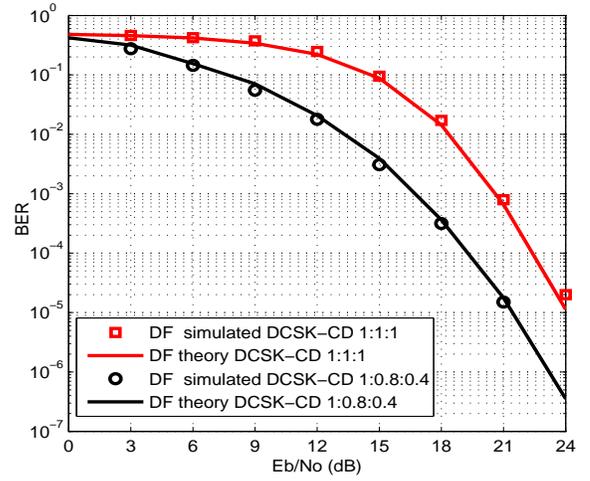}
\vspace{-0.3cm}
\caption{BER curves of the MIMO relay DCSK-CD systems with different distance ratios.}
\label{fig:Fig.9}
\end{figure}

To further justify the applicability of the proposed system, some practical environments are now discussed for this system with DF protocol. Assume that $d_{SD}: d_{SR}: d_{RD}=1: 1: 1$ and $2\beta = 128$. The BER curves for different $(M_R, M_D)$ pairs are shown in Fig.~\ref{fig:Fig.8}. One can observe that the BER performance of the proposed system will be improved as the diversity order $M_R M_D$ increases for
obtaining more spatial diversity gain. For example, the proposed system with $(M_R, M_D) = (3, 2)$ has a gain about $3$~dB over that with $(1, 1)$ at a BER of $2 \times 10^{-4}$, and it is likely to be even more in the higher SNR region. Also, the influence of different distance ratios on the system performance was investigated. The users always choose a relay which is much closer to the destination and cooperate with them to transmit messages. In this situation, we assume that $2\beta$ and $(M_R, M_D)$ here maintain $128$ and $(2, 2)$. The distance ratio $d_{SD}: d_{SR}: d_{RD}$ is set to be $1: 0.8: 0.4$ and $1:1:1$, respectively. The BER curves of these two cases with DF are plotted in Fig.~\ref{fig:Fig.9}. It illustrates that when $d_{SD}$ is fixed, the system with less $d_{SR}$ or $d_{RD}$ distance exhibits better performance for obtaining extra path-gain (less path-loss).
\begin{figure}[t]
\center
\includegraphics[width=3in,height=2.5in]{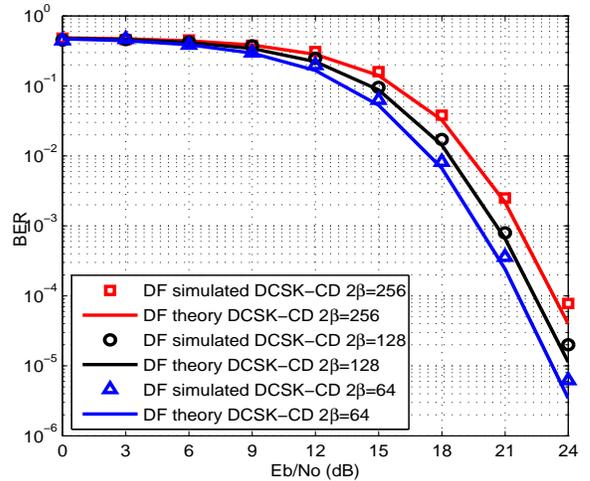}
\vspace{-0.3cm}
\caption{BER curves of the proposed systems with different spreading factors.}
\label{fig:Fig.10}
\end{figure}

In addition, Fig.~\ref{fig:Fig.10} depicts the BER curves of the systems with three
 different spreading factors. The parameters $d_{SD}: d_{SR}: d_{RD}, (M_R, M_D)$ and $m$ are assumed to be $1: 1: 1, (2, 2)$
 and $1$ while the global spreading factors $2\beta = 64, 128$ and $256$ are considered.
 In this figure, we observe that better performance can be expected for a smaller spreading factor when the energy per bit is constant
provided that $2\beta T_s \gg \tau_i$ since the average energy of each sampling point $E\{(c_i)^2\} = E_b /(2\beta)$
 will decrease as $2\beta$ becomes larger and thus results in worse anti-noise capacity.

\subsection{Discussion about the Threshold of Antenna Numbers for the Proposed System}
\label{subsect:threshold-DCSKCD}
Given a certain value of fading factor $m$ and path number $L$ of the Nakagami fading channel,
the diversity order of the proposed system is $M_R M_D$. As the energy per bit is set to constant,
i.e., $E_b$ is unit energy, the transmission energy of each individual channel constructed by
transmit-receive antenna in the R-D link is $E_b / (2 M_R M_D)$ (another $E_b/2$ energy is used by source in the $1^{\rm st}$ time slot).
 Consequently, the system performance should be improved as $M_R M_D$
increases when this parameter is small since the performance gain from a higher diversity order is larger than the performance loss
from the worse anti-noise capacity, especially in the high SNR region.
However, the system performance is no longer improved if $M_R M_D$ exceeds a certain threshold as the transmission
energy is extremely low so that the effect of the diversity gain can just offset the effect of the deteriorated anti-noise capacity.
Consequently,
there is a threshold of the value of $M_R M_D$ in the proposed system where the influence of parameter $M_R$ is equivalent to that of $M_D$,
and this threshold can be obtained through observation of the corresponding theoretical performance curves.

\begin{figure}[tbp]
\center
\includegraphics[width=3in,height=2.5in]{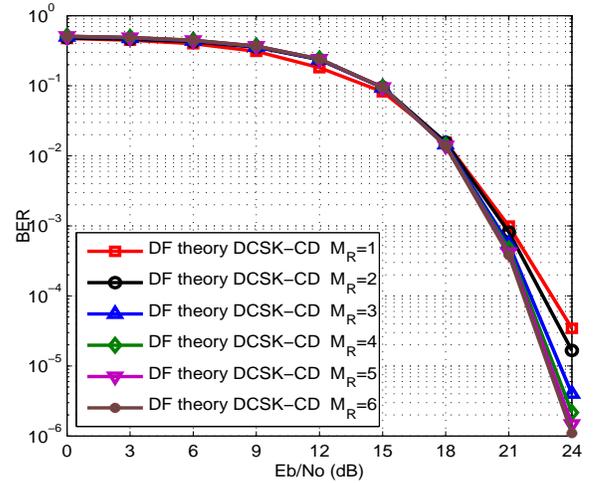}
\vspace{-0.3cm}
\caption{The BER curves of the proposed system with different numbers of relay antennas.
The number of destination antennas $(M_D)$ is set to $2$.
}
\label{fig:Fig.11}
\end{figure}

Based on the aforementioned analysis, the system performance will become better as the diversity order $M_R M_D$ becomes larger, but the complexity will increase simultaneously.
Moreover, the performance gain is negligible when $M_R M_D$ exceeds a certain threshold.
 So, a trade-off between system complexity and performance should be considered. As the effect of the parameters $M_R$ and $M_D$ are equivalent in such a system,
the discussion here is only focusing on the influence of the parameter $M_R$ while keeping $M_D$ constant, and vice versa. Assume that $d_{SD}: d_{SR}: d_{RD}=1: 1: 1, 2\beta = 128$ and $M_D = 2$.
 Fig.~\ref{fig:Fig.11} shows the exact BER curves obtained from numerical calculation with DF protocol in Sect.~\ref{sect:DCSKCD-BER}.
Referring to this figure, the system performance is almost no longer improved if the $M_R$ is
 over four, because the transmission energy for the channel that constructed by
 the transmit-receive antenna pair is extremely low, which makes the performance gain from the spatial
 diversity merely the same as the performance loss from the worse anti-noise capacity.
 Accordingly, $M_R^\ast=4$ and ${(M_R M_D)}^\ast=8$, can be seen as a threshold for $M_R$ and $M_R M_D$ in the proposed system, respectively.

\subsection{Performance Comparison between Approximate and Exact BER Results}
\label{subsect:Performance-APP}
From the numerical and simulation results shown in the previous subsections, we conclude that they agree quite well. In this subsection, the performance of the approximate (APP) BER expressions in Sect.~\ref{sect:DCSK-APPBER} is compared with the exact BER results in Sect. \ref{sect:exactconv_BER} and \ref{sect:DCSKCD-BER}, thereby verifying the accuracy of the closed-form expressions.

\begin{figure}[tbp]
\centering
\subfigure[]{ \label{fig:subfig:a} 
\includegraphics[width=3.0in,height=2.5in]{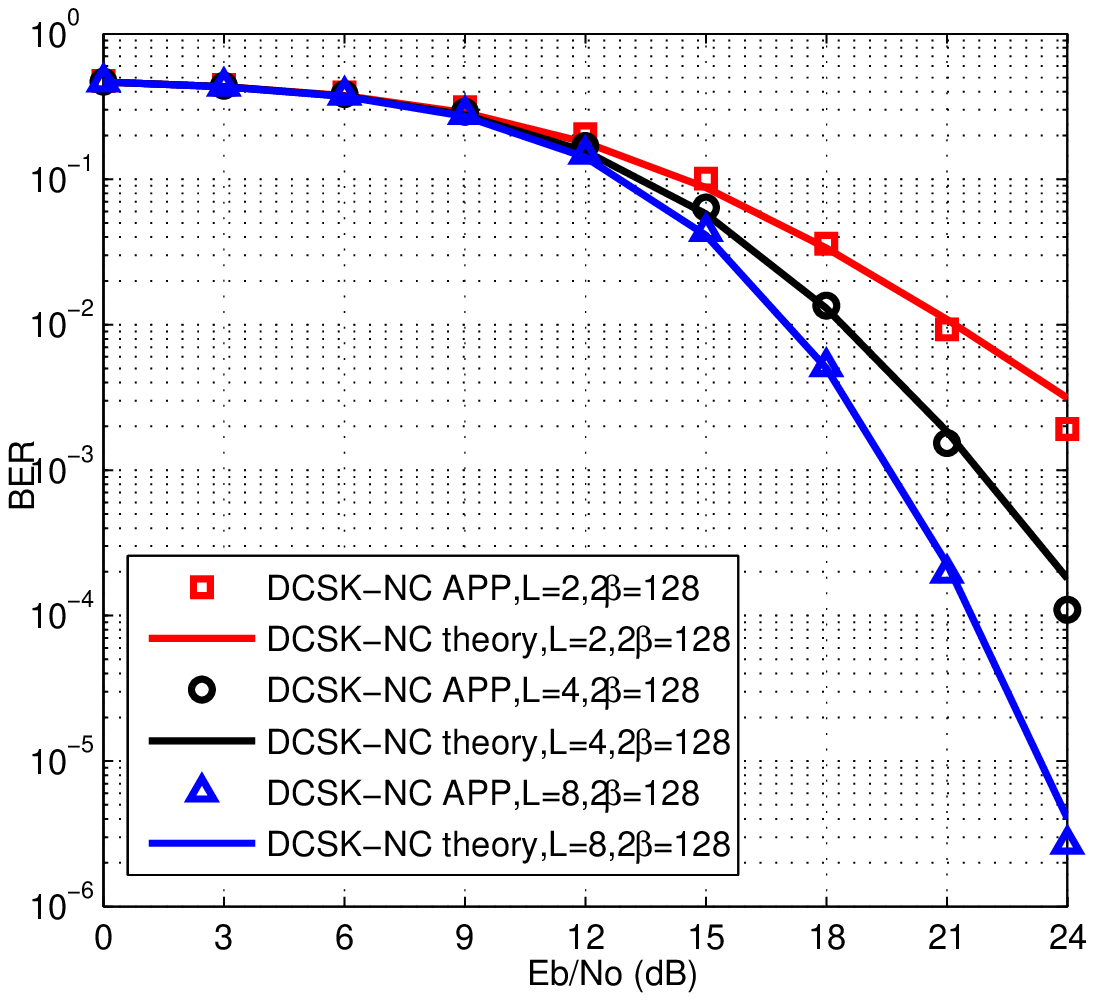}}
\subfigure[]{ \label{fig:subfig:b} 
\includegraphics[width=3.0in,height=2.5in]{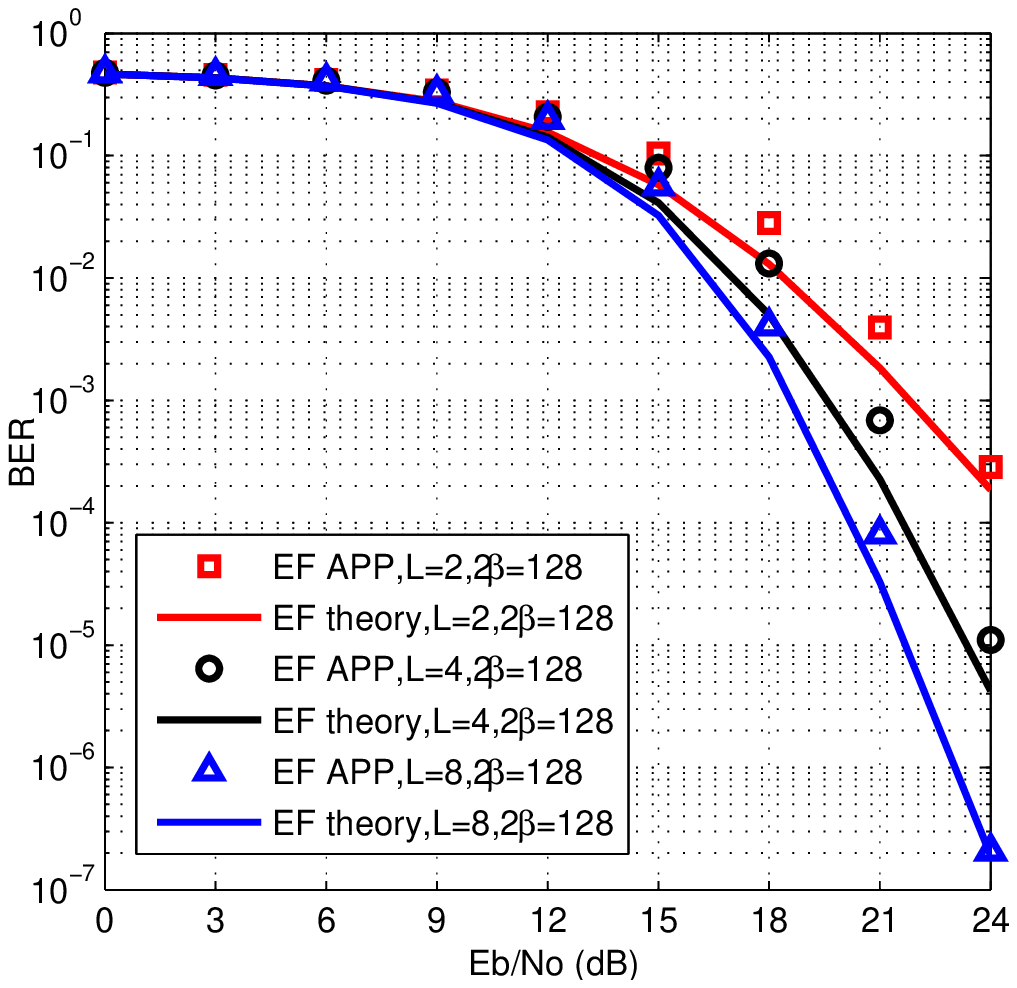}}
\vspace{-0.2cm}
\caption{The APP and the exact BER curves of the DCSK-NC system (a) and the MIMO relay DCSK-CD system (b) with different numbers of paths.}
\label{fig:Fig.12}  
\end{figure}

 We assume that the system parameters $d_{SD}:d_{SR}:d_{RD}, (M_R,M_D)$ and $2\beta$ are set to $1:1:1, (1,1)$ and $128$, respectively.
Therefore, the precondition of the formula \eqref{eq:DCSKCD-BER-APP} is fulfilled, i.e., $d_{SD}/d_{RD} = \sqrt{M_R}=1$. The APP BER and the exact BER curves of DCSK-NC and MIMO relay DCSK-CD systems with different numbers of paths are presented in Fig.~\ref{fig:Fig.12}. It implies that as the number of paths increases, the performance curves of approximate BER and exact BER expressions become more consistent for such two systems. In particular, referring to Fig.~\ref{fig:Fig.12}(b), the APP BER curve is much closer to the exact one when $L=8$ as compared to those of $L=2$ and $L=4$.
Based on the experimental results, we may conclude that the approximate BER expressions are expected to be more accurate with a large number of paths.
\vspace{-0.25cm}

\section{Conclusions} \label{sect:Conclusions}
In this paper, a MIMO relay DCSK-CD system has been proposed as a comprehensive cooperation scheme to strengthen the robustness of anti-fading in a wireless network, where both the relay and the destination can support multiple antennas, to get more spatial diversity gain. An exact BER formula and the corresponding closed-form approximate expression are derived over Nakagami fading channels, which agree well with the simulation results. Comparing to the conventional DCSK-NC and DCSK-CC systems, the performance gain of the proposed system with one antenna for relay and destination is about $4$~dB and $1$~dB at a BER of $4\times10^{-3}$, respectively, and more performance gain can be expected with a higher diversity order $M_R M_D$. Moreover, it was observed that a trade-off between system performance and complexity is needed when the energy of each user is kept constant, and an optimal value of the number of antennas at relay ($M_R^\ast=4$) is obtained through observation of the theoretical performance curves when $M_D$ is set to 2. According to these advantages, the proposed system has great potential in improving the performance of energy-constrained wireless networks, such as low-power and lower-cost WPANs.
In the future, we will strive to derive a close-form BER formula for the case of $d_{SD}/d_{RD} \neq \sqrt{M_R}$ thereby resolving the remainder practical problem in introducing the proposed system.

\bibliographystyle{IEEEtran}

\end{document}